# Electron spin resonance investigation of $Mn^{2+}$ ions and their dynamics in manganese doped $SrTiO_3$

V.V. Laguta,[1,3] I.V. Kondakova,[1] I.P. Bykov,[1] M.D. Glinchuk,[1] P.M. Vilarinho,[2] A. Tkach,[2] L. Jastrabik[3]

[1]*Institute for Problems of Material Science, NASc of Ukraine, Krjijanovskogo 3, 03680 Kiev, Ukraine*
[2]*Department of Ceramics and Glass Engineering,, University of Aveiro, 3810-193 Aveiro, Portugal*
[3]*Institute of Physics, AS CR, Cukrovarnicka 10, 16253 Prague, Czech Republic*

**Abstract**

Using electron spin resonance, lattice position and dynamic properties of $Mn^{2+}$ ions were studied in 0.5 and 2 % manganese doped $SrTiO_3$ ceramics prepared by conventional mixed oxide method. The measurements showed that $Mn^{2+}$ ions substitute preferably up to 97 % for Sr if the ceramics is prepared with a deficit of Sr ions. Motional narrowing of the $Mn^{2+}$ ESR spectrum was observed when temperature increases from 120 K to 240-250 K that was explained as a manifestation of off-center position of this ion at the Sr site. From the analysis of the ESR spectra the activation energy $E_a$ = 86 mV and frequency factor $1/\tau_0 \approx (2-10) \cdot 10^{-14}$ 1/s for jumping of the impurity between symmetrical off-center positions were determined. Both values are in agreement with those derived previously from dielectric relaxation. This proves the origin of dielectric anomalies in $SrTiO_3$:Mn as those produced by the reorientation dynamics of $Mn^{2+}$ dipoles.

## I. Introduction

Incipient ferroelectric SrTiO$_3$ (abbreviated as STO) is considered to be a classical displacive soft mode system where, however, the ferroelectric phase is suppressed by zero point quantum fluctuations of the soft mode leading to quantum paraelectricity down to the lowest temperatures, $T \to 0$ [1]. Owing to the unusually high polarizability, various types of polar phases (dipole glass, ferro-glass and ferroelectric) can be induced by dipole impurities [2] or by pressure [3] and by $^{18}$O isotope exchange [4,5]. From this point of view, SrTiO$_3$ is a promising material for many future applications of ferroelectrics, especially at cryogenic temperatures.

Among different types of doping impurities, the impurities possessing both the electric and magnetic dipole moments attract special attention of scientists because they allow to construct multiferroic materials that show unusual magneto-electric properties (see, for example [6]). Recent dielectric investigations of SrTiO$_3$ ceramic samples doped with Mn$^{2+}$ paramagnetic ions under special conditions (lacking Sr$^{2+}$ ions) [7] have shown strongly marked dielectric anomalies around 50 K: diffuse maximum in dielectric susceptibility and losses shifting to higher temperatures with increasing measurements frequency and amount of the manganese, which could indicate the onset of polar phase at $T \simeq 50K$ [8]. This study also points out that the anomalies can result from reorientation motion of dipoles formed in host matrix due to possible off-central position of Mn$^{2+}$ impurity ions substituting for the Sr$^{2+}$ ions. However, the dielectric measurements allow us to obtain only indirect evidence about origin of the dipoles responsible for the dielectric anomalies. To prove this idea and obtain more detailed information about the dipoles created by Mn impurity in the lattice, taking into account that Mn$^{2+}$ is a paramagnetic ion, it would be very fruitful to carry out ESR measurements which are the most direct method for determination of impurity-center structure and characteristics.

ESR studies of Mn impurities in SrTiO$_3$ have been performed previously by many authors. Because of multivalent character of Mn ions, several types of manganese centers were found and described both in single crystal [9,10] and ceramics [11]. Usually manganese dissolves into SrTiO$_3$ as Mn$^{4+}$(3d$^3$, S=3/2) ions which naturally substitute for Ti$^{4+}$ [9]. However, in reduced SrTiO$_3$ [10] and in ceramics, Mn$^{2+}$(3d$^5$, S=5/2) EPR spectra have been also observed. All these spectra were assigned to Mn$^{2+}$ substituted for Ti$^{4+}$ with (and without) local charge compensation. Due to very close ionic radii of Mn$^{2+}$ (0.08 nm) and Ti$^{4+}$ (0.10 nm), it can hardly be expected that for such substitution the manganese will be off-centrally shifted. Therefore, the origin of manganese dipole centers in SrTiO$_3$ is still not clear and calls for detailed investigation.

In the present work we report ESR measurements performed on several STO ceramic samples both nominally pure and doped by Mn (0.5 and 2 at. %) in which the dielectric anomalies were observed. Quantitative analysis of the measured Mn$^{2+}$ EPR spectra made possible to find out that manganese preferably substitutes for the Sr ions and create the reorientational dipoles originated from Mn$^{2+}$ displacement from the central position in the STO cubic lattice in several symmetrical directions. These reorientational Mn$^{2+}$ dipoles are responsible for the dielectric anomalies at $T \simeq 50 K$ (for Mn concentration 2 at %) as the dielectric relaxation and the relaxation dynamic of the Mn$^{2+}$ dipoles are described by approximately the same parameters.

II. Samples and experimental details

Manganese doped SrTiO$_3$ ceramic samples were prepared by the conventional mixed oxide method. Reagent grade SrCO$_3$, TiO$_2$ and MnO$_2$ were weighed according to the compositions Sr$_{1-x}$Mn$_x$TiO$_3$ with x=0.005 and 0.02 and small deficit of Sr in order to increase probability for Mn$^{2+}$ ions to occupy the Sr sites (for details, see Ref. 7). Room-temperature x-ray diffraction results indicated that all of the samples are of single cubic perovskite phase.

ESR measurements were performed at 9.5 - 9.6 GHz in the standard 3 cm wavelength range and at a temperature of 20-300 K. An Oxford Instrument ESR 900 cryosystem was used.

## III. Experimental results

Even undoped $SrTiO_3$ ceramic samples show intensive ESR spectrum of $Mn^{4+}$ impurity (Fig.1) which substitutes for $Ti^{4+}$. The spectrum was well described by a spin Hamiltonian with isotropic g-factor 1.9920±0.0002 and hyperfine constant $A = (71.3 \pm 0.1) \cdot 10^{-4} \ cm^{-1}$ (the $^{55}Mn$ isotope has nuclear spin I=5/2)) similar to those published by Muller et al. [9].

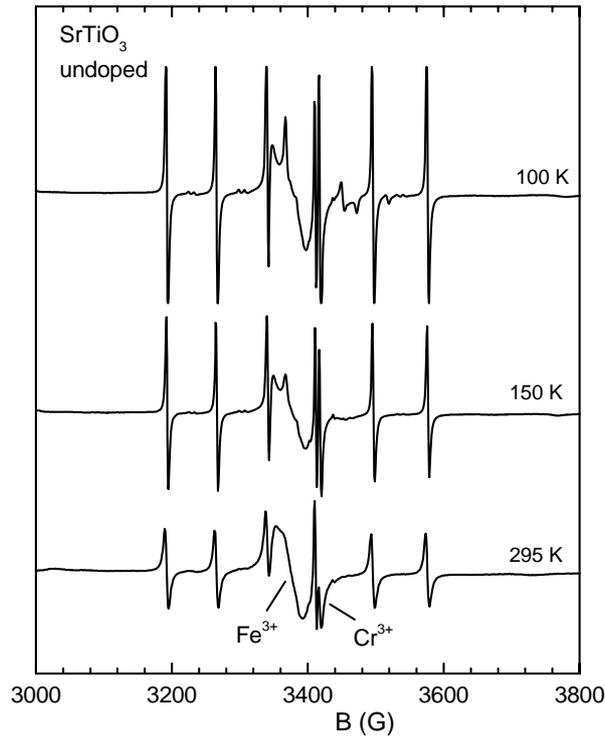

FIG.1: ESR spectra of undoped $SrTiO_3$ ceramics taken at several temperatures. Six narrow lines arise from $Mn^{4+}$ ions.

Besides the $Mn^{4+}$ impurity the samples also contained some amount of $Fe^{3+}$ and $Cr^{3+}$ ions (shown in Fig. 1, too), which both substitute for $Ti^{4+}$ as well. All these spectra show conventional behavior with temperature lowering, which is characteristic of the Ti cubic site, i.e. the increase of the intensity due to decrease of the line width and changing of the Boltzmann factor. Note that in single crystals of $SrTiO_3$ below the temperature of the structural phase transition $T_c \approx 105$ K, ESR spectra become slightly anisotropic (see, for example, Ref. [10]). However this anisotropy can hardly be observed in ceramics.

In SrTiO$_3$ ceramics specially doped with Mn$^{2+}$ ions along with Mn$^{4+}$ spectrum another spectrum arises that is also related to manganese ions. The new spectrum increases in intensity with increasing Mn concentration (Fig. 2 and Fig. 3). Note that the integral intensity of Mn$^{4+}$ spectrum also increases but not as much as the intensity of new spectrum, which we ascribe to Mn$^{2+}$ ions substituted for Sr$^{2+}$. The ratio of the integral intensities of Mn$^{2+}$ and Mn$^{4+}$ spectra is 4.5 and 35 for 0.5% and 2% Mn, respectively. The Mn$^{2+}$ spectrum at T > 100 K is also described by a spin Hamiltonian of the cubic symmetry with the following spectral parameters: $g = 2.0032 \pm 0.0002$ and $A = (82.8 \pm 0.2) \cdot 10^{-4}$ $cm^{-1}$. No fine structure splitting of the spectrum was observed at these temperatures.

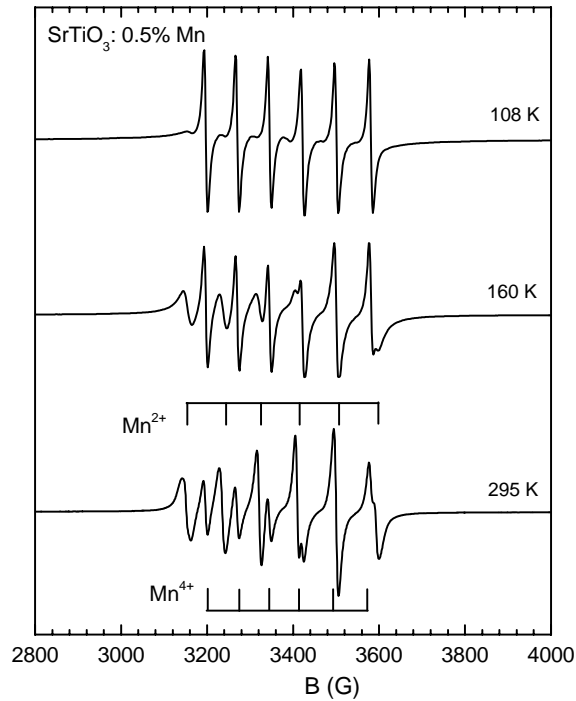

FIG. 2: ESR spectra of SrTiO$_3$: 0.5% Mn ceramics taken at several temperatures.

Unexpected feature of the Mn$^{2+}$ spectrum is that the width of each line of the Mn$^{2+}$ sextet nearly exponentially increases on lowering the temperature (Fig. 4). That is well seen in the spectrum with 2% Mn (Fig. 3), where the broadening at T<130 K is so large that the hyperfine (hf) lines become completely unresolved. At the same time one can see that the $Mn^{4+}$ linewidth did not change with temperature. The precise values of $Mn^{2+}$ linewidth were calculated performing deconvolution of the measured spectra into separated lines of $Mn^{2+}$ and $Mn^{4+}$ hf sextets, which is illustrated in Fig. 3 for the sample with 2% Mn. Note that at all temperatures the lineshape is Lorentzian. The temperature dependence of the linewidth, obtained in such a way, is shown in Fig. 4 for both concentrations of manganese. The residual linewidth at high temperatures mainly originates from the dipole-dipole interaction of $Mn^{2+}$ magnetic moments with each other and with $Mn^{4+}$. It is indeed larger in the sample with 2% Mn.

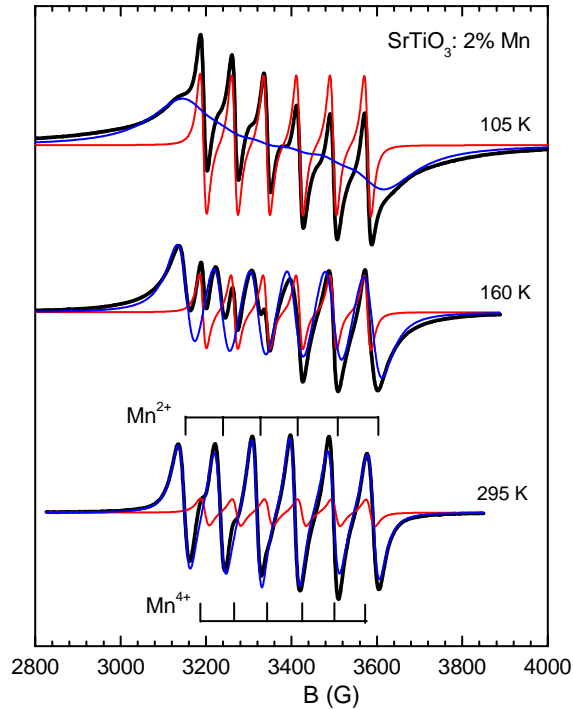

FIG. 3: (Color online) ESR spectra of $SrTiO_3$: 2% Mn ceramics taken at several temperatures (thick lines). Deconvolution of the experimental spectra into two sextets belonging to the $Mn^{2+}$ and $Mn^{4+}$ is shown by thin lines.

Such temperature behavior of the linewidth can indicate a motional narrowing of the spectrum, when impurity substitutes host ion in an off-center position and, thus, there can be fast jumping of the dipole between several symmetrically equivalent configurations leading to smearing of the fine structure splitting of the spectrum, which is expected for the non-cubic position of the impurity. In this sense, the observed behavior of the $Mn^{2+}$ spectrum is very similar to that in $KTaO_3$: Mn, where $Mn^{2+}$ is off-center substituted for $K^+$ ion (for details, see Ref. [12]).

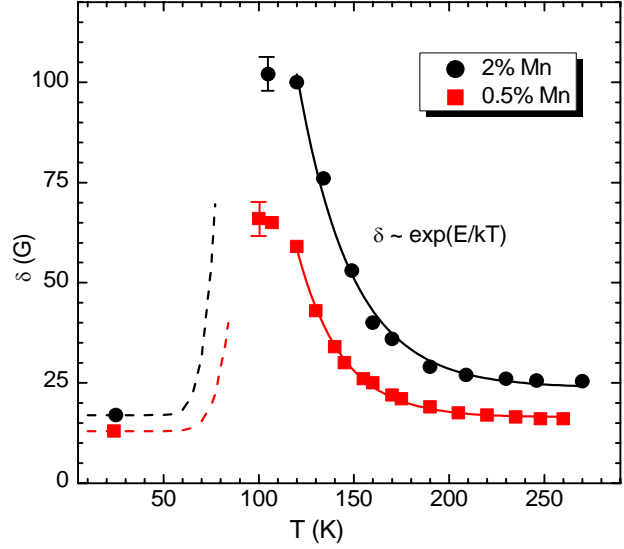

FIG. 4: (Color online) Temperature dependence of $Mn^{2+}$ $M_s = \pm 1/2$ transition linewidth (HWHM) for two manganese concentration: 0.5% and 2%. Linewidths at 25 K were derived from "static" powder spectra by using Eq. 3.

Because the measured spectral parameters, g-factor and hyperfine constant *A*, are typical for 2+ charge state of the manganese in perovskites (see, for example, [10,13,14],), there are no doubts that we are really dealing with $Mn^{2+}$ ions. Moreover, very similar spectra were observed previously in $SrTiO_3$ single crystal [10] and ceramics [11] doped with Mn. However, as it was pointed above, all these spectra were ascribed to $Mn^{2+}$ at $Ti^{4+}$ lattice sites. Therefore, a question arises if the $Mn^{2+}$ ions in our ceramic samples (and maybe in other mentioned studies) really substitute for $Sr^{2+}$ or they are occupied $Ti^{4+}$ octahedral sites.

Our assignment of the $Mn^{2+}$ spectrum as that belonging to $Mn^{2+}$ at $Sr^{2+}$ sites is supported by the following arguments:

(i) The specimens were synthesized with a deficit of Sr ions and therefore the manganese ions preferably occupy the Sr lattice sites (for details, see Ref. [7]);

(ii) Constant of the hyperfine splitting $A = 83 \cdot 10^{-4}$ $cm^{-1}$ is too large being for $Mn^{2+}$ at octahedral position, where the typical value is lower: for example, $A = 79.3 \cdot 10^{-4}$ $cm^{-1}$ in $BaTiO_3$ [14,15]. On the other hand, in $KTaO_3$ $Mn^{2+}$ at $K^+$ site has $A = 84 \cdot 10^{-4}$ $cm^{-1}$ [12], which is quite close to the hyperfine splitting measured by us for $Mn^{2+}$ in $SrTiO_3$. The Sr site is dodecahedrally coordinated by $O^{2-}$ ions, therefore the expected hyperfine splitting has to be larger then that for $Ti^{4+}$ octahedral site [16];

(iii) Large difference in ionic radii between the $Sr^{2+}$ and $Mn^{2+}$ [r(Sr) = 0.12 nm, r($Mn^{2+}$) = 0.08 nm) ] favours off-center position of the $Mn^{2+}$ ions and can explain the motional narrowing in its ESR spectrum. The difference in ionic radii between $Ti^{4+}$ and $Mn^{2+}$ is insufficient for appearance of $Mn^{2+}$ ions shift from the central position which is in agreement with sharp compression of unit cell observed earlier [7]. Dielectric experiments suggest the off-centrality of $Mn^{2+}$ ions too.

## IV. Analysis of the experimental data and discussion

The observed spectra transformation with temperature increasing from 110 to 250 K can be explained as a manifestation of dynamic process involving the $Mn^{2+}$ impurities. Theoretical description of EPR spectrum behavior in the presence of impurity hopping is given in many publications, see, for example, Refs. [17,18]. In our analysis we will use the results of publication [13], where the analysis of motional narrowing of EPR lines was performed for the case of S = 5/2 paramagnetic ion hopping between three magnetically inequivalent positions. This is adequate to our case when the paramagnetic off-center ion jumps near the central position.

We start from analysis of ESR spectra at low enough temperatures (T<<120 K), where the dynamic process is too slow to affect the line shape (so-called static regime). In the static regime $Mn^{2+}$ spectrum has to be described by the static spin Hamiltonian which contains zero field splitting (ZFS) terms. It can be presented in the form:

$$\hat{H} = g\beta\mathbf{B}\hat{\mathbf{S}} + A\cdot\hat{\mathbf{S}}\hat{\mathbf{I}} + D\left[\hat{S}_z^2 - \frac{1}{3}S(S+1)\right]. \tag{1}$$

Here we assumed the axial symmetry of the paramagnetic center (this is true at least to the temperature of $T_a \cong 105$ K). The axial crystal-field constant D reflects the local distortion of the lattice around $Mn^{2+}$ ion. In the present case, it is determined by the off-center shift of the ion, which can occur in several equivalent symmetrical directions so that the axial axis **Z** have different orientations relatively external magnetic field. This allows determining the direction of the ion displacement from single-crystal spectra. However, this cannot be done in ceramics or powders because their spectrum is averaged over all crystallites orientations. Moreover, we are not able to determine the ZFS constant D by conventional means, i.e. from positions of the fine $M_s \neq 1/2$ transitions because no fine structure is visible in $Mn^{2+}$ spectrum due to strong broadening of these transitions that usually occurs in ceramics. However, the value of D can be estimated from the shift of the central transition resonance field, which value for the conditions $|D|, |A| \ll g\beta B$ can be represented in the form [19]:

$$\begin{aligned}B_r(1/2 \leftrightarrow -1/2, m) = B_0(m) + \frac{2D^2}{B_0}(8\sin^2\theta - 9\sin^4\theta) \\ -2m\left[\frac{D^2 A}{B_0^2}(72\sin^2\theta - 73\sin^4\theta) + \frac{2DA^2}{B_0^2}(3\cos^2\theta - 1)\right] + ...,\end{aligned} \tag{2}$$

where $B_o(m)$ is the unperturbed resonance field of the m-th hyperfine transition, and $\theta$ is the angle between axial axis of the center and magnetic field. When $|D| > |A|$, we can keep in Eq. (2) only the first larger term with $D^2$. In this case the maximum shift of the resonance field is determined by the ratio $2D^2/B_0$, which we can use for crude approximation of the $Mn^{2+}$ hf linewidths at $T \simeq 100 - 120$ K. This leads to the following values of the constant D calculated from the data of Fig. 3: $280\cdot 10^{-4} cm^{-1}$ and $340\cdot 10^{-4} cm^{-1}$ for 0.5% and 2% of manganese, respectively.

More accurate determination of the constant D was carried out using splitting of the hyperfine lines which appears in the static spectrum at $T \leq 50-60\,K$. This splitting is well visible only for the left-hand (1/2,5/2 ↔ -1/2,5/2) transition (Fig. 5(a)) because other transitions are overlapped with strong $Mn^{4+}$ resonances. For simulation of the static spectrum we applied standard method by averaging the angular-dependent single-crystal spectra:

$$I(B) \propto \sum_{m=-5/2}^{m=5/2} \int_0^{\pi/2} \frac{dB_r(1/2,m)}{d(h\nu)} F\left(\frac{B_r(1/2,m)-B}{\delta}\right) \sin\theta\, d\theta, \qquad (3)$$

where $B_r(1/2,m)$ is defined by Eq. (2) and $F((B_r - B)/\delta)$ is the line shape function which was taken to be Lorentzian. Owing to the weak angular dependence of the central transition, the term $dB_r/d(h\nu)$ was treated as a constant. Contributions from the forbidden transitions were ignored too, due to their relatively small intensity when $D \ll g\beta B$.

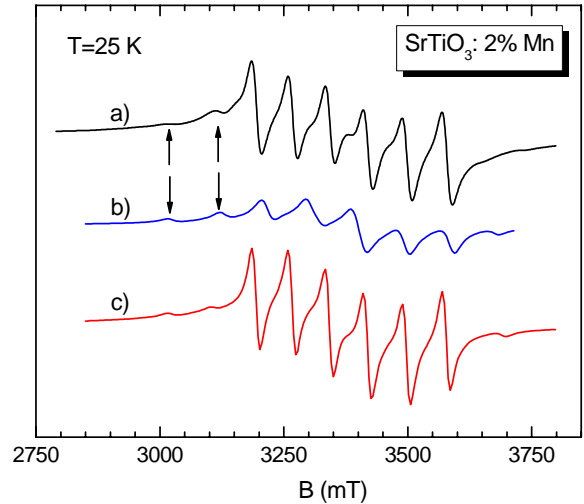

FIG. 5: (Color online) Experimental (a) and simulated (c) ESR spectra (central lines $M_S$= 1/2 ↔ $M_S$= -1/2) of SrTiO$_3$: 2% Mn in the static regime at 25 K. The spectrum (b) represents only simulated $Mn^{2+}$ resonances while the spectrum (c) allows also for $Mn^{4+}$ contribution. The splitting of the (1/2,5/2 ↔ -1/2,5/2) transition is shown by arrows.

The simulated powder spectra of the central-transition region are shown in Fig. 5 (b) and (c), respectively for only the single $Mn^{2+}$ and sum of $Mn^{2+}$ and $Mn^{4+}$ spectra. The best agreement between calculated and measured spectra were obtained taking for $Mn^{2+}$ $D=420 \cdot 10^{-4}$ $cm^{-1}$ and linewidth $\delta=2$ mT. The obtained value of the constant D seems as quite reasonable if we take into account that in $SrTiO_3$ the difference between ionic radii of $Sr^{2+}$ and $Mn^{2+}$ is smaller than that between $K^+$ and $Mn^{2+}$ in $KTaO_3$, where the D-constant is respectively larger, 0.148 $cm^{-1}$.

It should be noted that some contribution from the tetragonal distortion of the lattice in D-constant below the temperature 105 K seems to be present as well. However, our estimation of its value based on EPR data of $Gd^{3+}$ [20], which substitutes for $Sr^{2+}$ in the central position, shows that below the antiferrodistortive phase transition the D-constant can be changed in the value only by $(30-50) \cdot 10^{-4}$ $cm^{-1}$ that is much smaller with respect to the measured value $420 \cdot 10^{-4}$ $cm^{-1}$. We can thus neglect this mechanism in our analysis.

As the temperature is raised and, thus, the rate of the Mn jumps $1/\tau$ between off-center positions increases, the fine structure components, which could be detected in single crystal, start to broaden and finally completely disappear when the jumping rate exceeds the zero field splitting, $1/\tau > ZFS$. On further heating, all transitions merge around the central $M_S = -1/2 \leftrightarrow M_S = 1/2$ transition and give rise to a single isotropic spectrum, so that the motional averaging of the spectrum occurs. This occurs at temperatures 100-120 K (Fig. 4) and corresponds to the limit $ZFS < 1/\tau \leq \nu_0$, where $\nu_0$ is the microwave frequency. However, in ceramics the central transition spectrum being predominantly inhomogeneously broadened does not essentially "feel" this motion of the impurity because the motional broadening can not exceed the value $D^2/B_0$. Here, however, we can expect a change of the broadening mechanism from inhomogeneous to homogeneous at $1/\tau > ZFS$. As can be seen from Fig. 4, the width of hyperfine components begins to change only at T>120 K, where the motional narrowing phenomenon starts at the condition $1/\tau(T) > \nu_0 \sim 10^{10}$ $s^{-1}$. Obviously, in this fast jump limit (so-called relaxation limit), the description of ESR spectra in single crystals and ceramic samples becomes completely identical because all fine transitions merge with the central transition and the spectrum becomes completely isotropic. According to Ref. [13], in the relaxation regime the shape of each motionally narrowed hyperfine line is a single Lorentzian

$$I(B) = \frac{I_0 \cdot (\delta_r + \delta_0)}{[B - B_r]^2 + (\delta_r + \delta_0)^2}, \tag{4}$$

where $\delta_r$ is the motionally narrowed linewidth and $\delta_0 \simeq 1.8\,mT$ is the residual linewidth. The linewidth $\delta_r$ for S = 5/2 ion can be written [13] as

$$\delta_r \simeq 2.8 D^2 \tau, \tag{5}$$

where τ is the relaxation time of ion hopping. This formula was used by us for the extraction of the relaxation time τ from the temperature dependence of narrowed EPR spectra lines. The result is shown in Fig. 5 as a dependence of the relaxation rate 1/τ *vs* the reciprocal temperature.

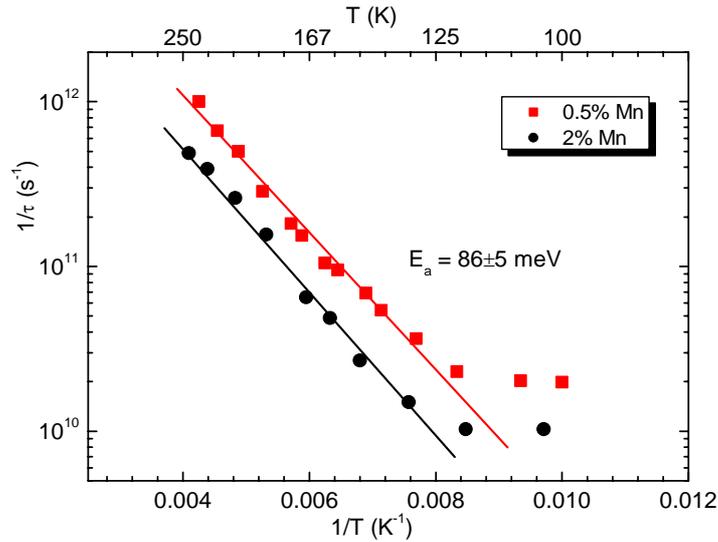

FIG. 6: (Color online) The hopping rate of $Mn^{2+}$ dipoles in $SrTiO_3$ as function of the reciprocal absolute temperature for two manganese concentrations: 0.5% and 2%. The points were derived from the experimental spectra as described in the text. Slopes of the straight lines correspond to the activation energy E = 86 meV of $Mn^{2+}$ hopping between off-center positions in $SrTiO_3$ lattice (see Eq. 6).

One can see that the temperature dependence of the jumping rate of impurity well follows the exponential law and can be described by an Arrhenius behavior

$$\frac{1}{\tau} = \frac{1}{\tau_0}\exp(-E_a/kT) \tag{6}$$

in both 0.5% and 2% Mn doped samples with the following kinetic parameters: the barrier height $E_a = 86\pm5$ meV, relaxation time pre-factor $\tau_0 \approx 1.0\times10^{-13}$ s and $\tau_0 \approx 1.8\times10^{-14}$ s for 0.5% and 2% Mn concentration, respectively.

The comparison of these values with the kinetic parameters obtained from the dielectric measurement data [8] shows good coincidence at least for the sample with 2% Mn. For lower Mn concentration the barrier height calculated from dielectric relaxation gradually decreases down to 34 mV at doping level of 0.25 %. In our opinion, such difference of ESR and dielectric measurements data are related to the fact that at low Mn concentration other relaxation mechanisms and sources can dominate in the dielectric response. This is also confirmed by the fact that the fraction of $Mn^{2+}$ ions is much less for the lower concentration of manganese and maybe comparable with concentration of any other defects in the sample. On the other hand, in case of ESR only relaxation dynamics of $Mn^{2+}$ ions are measured. Moreover, because homogeneous linewidth $\delta_r$ is directly proportional to the relaxation time $\tau$, the calculated barrier height is not sensitive to the model of the ZFS constants calculation. Small difference of the relaxation time pre-exponent value with the change of Mn concentration obtained from the description of the temperature dependence of linewidth can be related to the interaction between dipole moments. Similar behavior was observed in the dielectric measurements [8].

In conclusion, we point out that though our study was not able to find out the direction of the $Mn^{2+}$ off-center displacement one can assume with high probability that is occurs at simple <100> cubic directions. This follows from the fact that in another very similar quantum paraelectric $KTaO_3$ all known dipole impurities situated at K site (for example, $Li^+$, $Na^+$, $Mn^{2+}$ [2]) are always shifted along <100> axes. Obviously, this problem can be definitively resolved by measuring ESR or NMR spectra of SrTiO$_3$: Mn$^{2+}$ single crystal.

The good agreement obtained between the data of the dielectric and EPR investigations of the SrTiO$_3$ ceramic samples provides a convincing evidence for the suggested model of the dynamic process involving random jumps of Mn$^{2+}$ off-center ions.

**Acknowledgement**

This work was supported by the project 1M06002 of the MSMT CR.